\input harvmac
\lref\wen {X.-G. Wen, "Origin of Light",
hep-th/0109120.}
\lref\bfss {T. Banks, W.Fischler, S.H. Shenker, L. Susskind,
Phys. Rev. D55 (5112) 1997, hep-th/9610043.}
\lref\taylor {W. Taylor,
"M(atrix) Theory: Matrix Quantum Mechanics as a Fundamental Theory",
hep-th/0101126.}
\lref\dounglas {M.R. Dounglas, JHEP 9797 (004) 1997,
hep-th/9612126.}
\lref\bss {T. Banks, N. Seiberg, S. Shenker,
Nucl. Phys. B490 (91) 1997, hep-th/9612157.}

\def\<{\langle}
\def\>{\rangle}

\Title{hep-th/0110030}
{\vbox{\centerline{Gauge Field on Brane in M-atrix Theory}}}
\smallskip
\centerline{Takeshi Inagaki}
\smallskip
\centerline{\it Yamato Laboratory, IBM Japan}
\smallskip
\centerline{\it Shimotsuruma, Yamato-shi, Kanagawa, Japan}

\bigskip
\bigskip

\noindent
In this ultra short note,
gauge field propagation in D-brane configuration of M theory
in the BFSS matrix formulation is considered. Noncommutativity
of the space plays a key role for appearance of gauge fields
as physical degrees of freedom.

\vskip 3cm
\noindent
\Date{Oct, 2001}

\newsec{ Introduction}
The BFSS matrix theory \refs{\bfss} is conjectured as
a second quantized description of M theory.
Among many features of this theory,
(see \refs{\taylor} for review),
appearance of a D-brane configuration \refs{\bss} from
the matrix algebra is very important for understanding
underlying M theory.
Recently X.-G. Wen \refs{\wen} proposed a mechanism in which
a $U(1)$ gauge field excitation appears in a $SU(N)$ spin model
on a 3D cubic lattice. He suggested existence of gapless
excitation is ensured by a symmetry group named quantum orders
in a system. 

In this note, we adapt this argument to the BFSS model
and see how a gauge field propagates on a D-brane.
Gauge degree of freedom is introduced in the model in a rather
trivial way and it dose not couple to other physical degrees of
freedom locally.
This model can be regarded as a kind of topological model
where gauge transformation appears through the global structure of
a base manifold. When the space is non-commutative, a close loop is
not contractive to a point and gauge fields become to propagate
along it. 

\newsec{ Origin of light in the BFSS matrix model}
The interaction term of bosonic fields in the BFSS model is
$$
H_I=tr \sum_{a<b} [X^a, X^b]^2
$$
where $a,b=1,2,.,9$ are indexes for space coordinates and $X_a$
is a $NxN$ matrix with $N \rightarrow \infty$.
Formally, this Hamiltonian can be decomposed in terms of more
fundamental interaction by introducing new fermionic variables
$\psi_l$,
$$
H_0=\sum_a \psi_l^* X_a^{lm} \psi_m
$$
where $\{\psi_l^*, \psi_m\}=\delta_{l,m}$ and $l,m=1,.,N$.
This reformation enlarges the Hilbert space and brings extra
states in the theory which should be projected out.
Among Feyman diagrams, graphs which form a square in a plane
are allowed after this projection. We can introduce
$U(1)$ gauge transformation acts on $\psi_l$ as
$$
\psi_l \rightarrow e^{i\theta_l} \psi_l.
$$
Original BFSS Hamiltonian is invariant under this gauge
transformation without any gauge field.
This gauge transformation acts on $X_a$ as
$$
X_a^{lm} \rightarrow e^{-i\theta_l} X_a^{lm} e^{i\theta_m}.
$$
In addition, we can introduce a gauge field $A_a^{lm}$ as
$$
X_a^{lm} e^{i(A_a^{lm})-\theta_l+\theta_m}.
$$
Now consider a square diagram in a plane in $a$-$b$ direction.
If the loop can be contractive to a point, it must be
$$
A_a^{ij}+A_b^{jk}+A_a^{kl}+A_b^{li}=0
$$
that means $A_{a,b}$ is pure gauge and it includes
no physical degrees of freedom. 
On the other hand, if it is not contractive, it can be
$$
A_a^{ij}+A_b^{jk}+A_a^{kl}+A_b^{li}=2\pi n
$$
for an arbitrary integer $n$ and there exists
physical degrees in the gauge field.
This suggests existence of non-contractive loop
diagrams is a key factor for existence of gauge
symmetry which couples to physical states. 
To check that this gauge fields carry energy,
consider momentum propagation in a loop diagram.
Momentum propagates along edges $C$ of this square is
$$
kl \psi=\int_C i{d\over{dx}}\psi
={i\over2}(X_a X_b X_a X_b - X_b X_a X_b X_a)\psi
$$
where $l$ is length of the path.
This momentum $k$ vanishes at the absence of D-brane
$[X_a, X_b]=0$. In the D2 brane configuration $[X_a,X_b]=iI$,
$k$ equals ${1\over l}$.
Along a long path, excitation with small momentum propagates
and this means the gauge field is massless.

\newsec{ Discussion}
As pointed out by Wen, appearance of gauge fields and their mass
(gauge symmetry breaking) will be determined from
symmetric structure (quantum order) in $X_a$. So more detailed
analysis in this direction will reveal relations among gauge
symmetries and the BFSS matrix theory in an apparent way.
Especially, matrix theory in orbifolds \refs{\dounglas}
seems to be an accessible example.

\listrefs
\end